\documentclass[10pt]{extarticle}
\usepackage{graphicx}
\usepackage{xcolor}
\usepackage{comment}
\usepackage{amssymb,amsmath}
\def\beq{\begin{eqnarray}}
\def\eeq{\end{eqnarray}}
\usepackage[titletoc,toc,title]{appendix}
\usepackage{hyperref}
\usepackage{amsfonts}

\begin{document}

\title{Spectral sum rules on a $d$--sphere}
\author{Paolo Amore \\
\small Facultad de Ciencias, CUICBAS, Universidad de Colima,\\
\small Bernal D\'{i}az del Castillo 340, Colima, Colima, Mexico \\
\small paolo@ucol.mx}

\maketitle

\begin{abstract}
We derive spectral sum rules for inverse powers of the eigenvalues of the Helmholtz equation on a $d$-sphere in the presence of an arbitrary density. By adopting a rigorous renormalization scheme, we remove the divergent contributions of the zero mode and obtain exact expressions for the sum rules without requiring an explicit determination of the eigenvalues, which is generally impossible. As an application, we derive explicit sum rules for the density $\Sigma(\Omega) = 1 + \kappa Y_{1,\vec{0}}(\Omega)$ in $d=3,4,5$ dimensions and compare them with  numerical estimates obtained by approximating the low-lying part of the spectrum with the Rayleigh--Ritz method and the high-energy part with Weyl's formula.
\end{abstract}

\section{Introduction}
\label{sec:intro}

We consider the weighted Laplacian eigenvalue problem on the unit $d$-sphere,
\begin{equation}
-\Delta_{S^{d}} \psi_n(\Omega_d) = E_n \Sigma(\Omega_d)\psi_n(\Omega_d)\,,
\label{eq_Helmoltz_dens}
\end{equation}
where $\Omega_d$ denotes the angular coordinates on $S^d$ and
\[
\Delta_{S^{d}} =
\frac{1}{\sin^{d-1}\theta_1}\frac{\partial}{\partial \theta_1}
\left(
\sin^{d-1}\theta_1 \frac{\partial}{\partial \theta_1}
\right)
+\frac{1}{\sin^2\theta_1}\,\Delta_{S^{d-1}}
\]
is the Laplace--Beltrami operator on the sphere.\footnote{Throughout the paper, $d$ denotes the dimension of the sphere, while $D=d+1$ is the dimension of the ambient Euclidean space.} For $d=2$, Eq.~\eqref{eq_Helmoltz_dens} describes the normal modes of a spherical membrane with variable density $\Sigma$.

Equations of the form  (\ref{eq_Helmoltz_dens}), although not generally posed on spheres, arise in several physical settings, including optics, where they govern wave propagation in media with spatially varying refractive index~\cite{Leonhardt2009PerfectImaging}; geophysics, where they model surface-wave propagation in laterally heterogeneous media~\cite{TrompDahlen1993PotentialRepresentation}; and quantum mechanics, in effective descriptions with position-dependent mass~\cite{LevyLeblond1995PDM}.

Our interest lies in the spectral sum rules
\[
Z_p \equiv \sum\nolimits_n' \frac{1}{E_n^p}\,,
\qquad p=2,3,\dots,
\]
where the prime indicates that the zero mode is omitted. In  Ref.~\cite{Amore20}, we analyzed these sum rules for the Helmholtz equation on $S^2$ with non-uniform density. In general, a direct evaluation of $Z_p$ would require explicit knowledge of the spectrum $\{E_n\}$, which is available only in special cases such as uniform density. For generic densities one may appeal to perturbation theory when the inhomogeneity is weak, but this yields only approximate expressions and does not provide exact results for arbitrary media.

The key observation of Ref.~\cite{Amore20} is that the sum rules may be rewritten as traces of suitable operators. Since the trace is basis independent, one may evaluate it in any complete orthonormal basis on the sphere, without requiring the exact eigenfunctions of the weighted problem. This makes the basis of the uniform-density problem, namely the spherical harmonics, particularly convenient.

However, this reformulation introduces a subtle point. In the exact eigenbasis, the zero mode can simply be excluded from the spectral sum. By contrast, when the trace is evaluated in the spherical-harmonic basis, the contribution associated with the zero mode appears as a divergent term and must therefore be removed by a suitable renormalization procedure. In Ref.~\cite{Amore20} we showed how this can be done explicitly on $S^2$, obtaining general exact expressions for the corresponding spectral sum rules for arbitrary densities.

The purpose of the present paper is to extend that construction to the unit $d$-sphere, with $d=3,4,\dots$, by working in the complete orthonormal basis of hyperspherical harmonics. In this way, we generalize the renormalized trace formulation of Ref.~\cite{Amore20} to higher dimensions and derive exact expressions for the spectral sum rules of the weighted problem on $S^d$.

Spectral sum rules have also been studied extensively in other settings. Itzykson, Moussa, and Luck~\cite{Itzykson86} derived closed integral expressions for sum rules involving inverse powers of the Dirichlet spectrum of the planar Laplacian on arbitrary two-dimensional domains. Their method relies on conformally mapping the problem to the unit disk, thereby avoiding any explicit determination of the eigenvalues. Using the same strategy, Berry~\cite{Berry86} extended the analysis to Aharonov--Bohm quantum billiards and obtained explicit results for several billiard geometries. Related questions for the circular Aharonov--Bohm billiard were investigated by Steiner~\cite{Steiner87}, building on earlier techniques developed for confinement potentials in Ref.~\cite{Steiner85}.

Subsequent work broadened the range of admissible geometries and boundary conditions. Kvitsinsky~\cite{Kvitsinsky96} studied sum rules for domains that are small deformations of the disk, with particular attention to regular $N$-gons. For simply connected planar regions, Dittmar~\cite{Dittmar02} derived sum rules for both fixed and free membrane problems through conformal mapping to the unit disk, and additional domain-specific evaluations were later reported in Ref.~\cite{Dittmar11}. In a different direction, Dostani\'c~\cite{Dostanic11} computed the regularized trace of the inverse Dirichlet Laplacian for bounded convex domains.

In our earlier work~\cite{Amore13A,Amore13B,Amore14,Amore18}, we developed general integral representations for spectral sum rules of inhomogeneous strings and membranes under several boundary conditions. The Neumann and periodic cases considered in Ref.~\cite{Amore14} are especially delicate because the presence of a zero mode makes the corresponding traces singular unless an appropriate prescription is adopted. A systematic regularized sum rule was later introduced in Ref.~\cite{Amore18}, either by exploiting symmetry or by relating problems with different boundary conditions. The renormalization procedure used in Ref.~\cite{Amore20} for the $2$-sphere is a direct implementation of the approach first developed in Ref.~\cite{Amore14}.

The paper is organized as follows. In Section~\ref{sec:general} we review the properties of hyperspherical harmonics needed in the analysis. In Section~\ref{sec:renorm} we express the sum rules as traces of appropriate operators and discuss their renormalization. In Section~\ref{sec:application} we apply the general formalism to the density
\[
\Sigma(\Omega)=1+\kappa Y_{1,\vec{0}}(\Omega)
\]
in $d=3,4,5$ dimensions. Finally, Section~\ref{conclusions} contains our conclusions.

\section{General properties}
\label{sec:general}

In this section we review some basic properties of the eigenvalues and eigenfunctions of Eq.~\eqref{eq_Helmoltz_dens} in the case of constant density \(\Sigma\). In this limit, the equation reduces to
\begin{equation}
-\Delta_{S^{d}}  Y_{\ell,\vec{m}}(\Omega_d)   = \lambda_\ell^{(d)}\, Y_{\ell,\vec{m}}(\Omega_d) \ ,
\end{equation}
where
\begin{equation}
\lambda^{(d)}_\ell = \ell (\ell + d-1)\,,
\qquad \ell = 0,1,\dots,
\end{equation}
are the eigenvalues of the negative Laplacian on the \(d\)-sphere, with degeneracy
\begin{equation}
g_\ell^{(d)} = \frac{(2\ell + d-1)(\ell + d-2)!}{\ell!\,(d-1)!}\ .
\end{equation}
The functions \(Y_{\ell,\vec m}(\Omega_d)\) are the hyperspherical harmonics~\cite{Avery85}, which reduce to the ordinary spherical harmonics for \(d=2\). They are labeled by the quantum numbers \(\ell=m_1\) and \(\vec m=(m_2,\dots,m_d)\), subject to
\[
m_1=\ell\in\mathbb N_0,
\qquad
\ell=m_1\ge m_2\ge\cdots\ge m_{d-1}\ge |m_d| \ .
\]

Using these eigenvalues, one may define the sum rules
\[
\zeta^{(d)}(p) = \sum_{\ell =1}^\infty \frac{g_\ell^{(d)}}{\left(\lambda_\ell^{(d)}\right)^p}\,,
\qquad
p=p_{\rm min}(d),\,p_{\rm min}(d)+1,\dots,
\]
where \(p_{\rm min}(d)\) is the smallest integer for which the series converges in \(d\) dimensions:
\[
p_{\rm min}(d)=\left\lfloor \frac{d+2}{2}\right\rfloor
=
\left\{
\begin{array}{ccc}
\dfrac{d+2}{2} & , & d \ \text{even} \\[4pt]
\dfrac{d+1}{2} & , & d \ \text{odd}
\end{array}
\right. .
\]

In particular,
\[
\begin{aligned}
\zeta^{(2)}(2) &= 1 \\
\zeta^{(3)}(2) &= \frac{1}{16}+\frac{\pi^2}{12} \\
\zeta^{(4)}(3) &= \frac{2 \zeta (3)}{27}+\frac{23}{1458} \\
\zeta^{(5)}(3) &= \frac{5}{6144}+\frac{19 \pi ^2}{2304}\\
&\dots
\end{aligned}
\]

The evaluation of \(\zeta^{(d)}(p)\) is straightforward in the homogeneous case, since the eigenvalues are known explicitly and the zero mode can simply be excluded from the sum.

The Green's function of the Laplacian on \(S^d\) reads~\cite{Szmytkowski06,Szmytkowski07}
\[
\mathcal{G}^{(d)}(\eta, {\bf n}, {\bf n}')
=
\frac{\pi}{(d-1)\,\mathrm{Vol}(S^{d})\,\sin \!\bigl(\pi \lambda(\eta)\bigr)}
\, C_{\lambda(\eta)}^{((d-1)/2)}(-{\bf n}\cdot{\bf n}') \ ,
\]
where
\[
\mathrm{Vol}(S^{d})= \frac{2\pi^{\frac{d+1}{2}}}{\Gamma\!\left(\frac{d+1}{2}\right)}
\]
is the area of the \(d\)-sphere, \(C_\lambda^{(\alpha)}(x)\) is the Gegenbauer function of the first kind, and \({\bf n}\) and \({\bf n}'\) are unit vectors on \(S^d\). Here
\begin{equation}
\lambda(\eta)=\frac{1}{2}\left(\sqrt{(d-1)^2+4\eta}-d+1\right)\ .
\end{equation}

This Green's function satisfies
\[
\left( \Delta_{S^{d}} + \eta \right)\mathcal{G}^{(d)}(\eta, {\bf n}, {\bf n}')
=
\delta(\Omega_d-\Omega_d')
\]
and admits the spectral representation
\[
\mathcal{G}^{(d)}(\eta, {\bf n}, {\bf n}')
=
\sum_{\ell=0}^{\infty}
\sum_{\vec m}
\frac{
Y_{\ell,\vec m}(\Omega)\,
Y^*_{\ell,\vec m}(\Omega')
}{
-\ell(\ell+d-1)+\eta
}\ ,
\]
where \(\ell \ge m_2 \ge \cdots \ge m_{d-1} \ge |m_d|\) for \(d\ge 3\). To simplify the notation, from now on we write \(\Omega\) instead of \(\Omega_d\).

It is convenient to introduce the higher-order Green's functions
\begin{equation}
\mathcal{G}^{(d,p)}(\eta, {\bf n}, {\bf n}')
\equiv
\frac{(-1)^p}{p!}\,
\frac{\partial^p}{\partial \eta^p}
\mathcal{G}^{(d)}(\eta, {\bf n}, {\bf n}') \ ,
\end{equation}
which have the spectral representation
\[
\mathcal{G}^{(d,p)}(\eta, {\bf n}, {\bf n}')
=
\sum_{\ell=0}^{\infty}
\sum_{\vec m}
\frac{
Y_{\ell,\vec m}(\Omega)\,
Y^*_{\ell,\vec m}(\Omega')
}{
\left(-\ell(\ell+d-1)+\eta \right)^{p+1}
}\ .
\]

Rather than working directly with \(\mathcal{G}^{(d)}(\eta,{\bf n},{\bf n}')\), we introduce the Green's function \(G^{(d)}(\Omega,\Omega')\) satisfying
\[
-\Delta_{S^{d}}\, G^{(d)}(\Omega,\Omega')
=
-\frac{1}{\mathrm{Vol}(S^{d})}+\delta(\Omega-\Omega') \ ,
\]
whose spectral representation is
\[
G^{(d)}(\Omega,\Omega')
=
\sum_{\ell=1}^{\infty}
\frac{1}{\ell(\ell+d-1)}
\sum_{\vec m}
Y_{\ell,\vec m}(\Omega)\,
Y^*_{\ell,\vec m}(\Omega') \ .
\]

Similarly, we define the higher-order Green's functions
\[
G^{(d,q)}(\Omega,\Omega')
\equiv
\sum_{\ell=1}^{\infty}
\frac{1}{\left(\ell(\ell+d-1)\right)^{q+1}}
\sum_{\vec m}
Y_{\ell,\vec m}(\Omega)\,
Y^*_{\ell,\vec m}(\Omega') \ ,
\]
which satisfy
\[
-\Delta_{S^{d}}\, G^{(d,q)}(\Omega,\Omega')
=
G^{(d,q-1)}(\Omega,\Omega') \ .
\]
These functions will be used in the next section to derive the spectral sum rules.

The hyperspherical harmonics satisfy the orthogonality relation
\[
\int_{S^{d}}
Y_{\ell,m_1,\ldots,m_{d-1}}(\Omega)\,
Y^{*}_{\ell',m_1',\ldots,m_{d-1}'}(\Omega)\,
\mathrm{d}\Omega
=
\delta_{\ell\ell'}\,
\prod_{i=1}^{d-1}\delta_{m_i m_i'}
\]
and the completeness relation
\[
\sum_{\ell=0}^{\infty}
\sum_{\vec m}
Y_{\ell,\vec m}(\Omega)\,
Y^{*}_{\ell,\vec m}(\Omega')
=
\delta^{(d)}(\Omega,\Omega') \ .
\]

Using orthogonality, one immediately finds
\[
G^{(d,p+q-1)}(\Omega,\Omega')
=
\int_{S^d}
G^{(d,p)}(\Omega,\Omega'')\,
G^{(d,q)}(\Omega'',\Omega')\,
\mathrm{d}\Omega'' \ ,
\qquad p,q=0,1,2,\dots
\]

The addition theorem for the hyperspherical harmonics is~\cite{Avery85}
\[
\sum_{\vec m}
Y_{\ell,\vec m}(\Omega)\, Y^{*}_{\ell,\vec m}(\Omega')
=
\frac{g_\ell^{(d)}}{\mathrm{Vol}(S^{d})}\,
\frac{
C_\ell^{\left(\frac{d-1}{2}\right)} \!\left(\Omega\cdot\Omega'\right)
}{
C_\ell^{\left(\frac{d-1}{2}\right)}(1)
} \ .
\]

An explicit representation of the hyperspherical harmonics on \(S^d\) is
\[
Y_{\ell, m_2, \ldots, m_{d}}(\Omega)
=
(-1)^{\frac{m_d + |m_d|}{2}} \,
\mathcal N_{\ell,\mathbf m}\, e^{\,i m_{d}\phi}
\prod_{k=1}^{d-1}
\left(\sin\theta_k\right)^{m_{k+1}}
C_{\,m_k-m_{k+1}}^{\,m_{k+1}+\frac{d-k}{2}}
\!\left(\cos\theta_k\right),
\]
where \(\Omega=(\theta_1,\ldots,\theta_{d-1},\phi)\).

For \(k=1,\ldots,d-1\), we define
\[
n_k:=m_k-m_{k+1}\in\mathbb N_0,
\qquad
\lambda_k:=m_{k+1}+\frac{d-k}{2},
\]
so that the normalization constant may be written as
\[
\mathcal N_{\ell,\mathbf m}
=
\left[
2\pi
\prod_{k=1}^{d-1}
\left(
\pi\,2^{\,1-2\lambda_k}\,
\frac{
\Gamma(n_k+2\lambda_k)
}{
n_k!\,(n_k+\lambda_k)\,\Gamma(\lambda_k)^2
}
\right)
\right]^{-1/2}.
\]

Finally, another property that will be useful later is
\begin{equation}
\int_{S^d}
Y_{\ell_1, \vec{m}_1}(\Omega)\,
Y_{\ell_2, \vec{m}_2}(\Omega)\,
Y_{\ell_3, \vec{m}_3}(\Omega)\,
\mathrm{d}\Omega
=
\left(
\begin{array}{ccc}
\ell_1  & \ell_2 & \ell_3 \\
\vec{m}_1 & \vec{m}_2 & \vec{m}_3
\end{array}
\right) ,
\end{equation}
which generalizes the Wigner \(3j\) symbols. An explicit expression for these coefficients is given in Ref.~\cite{Avery85}. The corresponding selection rules include
\begin{equation}
\begin{aligned}
|\ell_1-\ell_2| \le \ell_3 \le \ell_1+\ell_2,
\qquad
\ell_1+\ell_2+\ell_3 \in \{2k \mid k\in\mathbb N_0\}, \\
m_{d,1}+m_{d,2}+m_{d,3}=0 \ .
\end{aligned}
\end{equation}

\section{Renormalization}
\label{sec:renorm}

We now describe the procedure that allows one to calculate the corresponding sum rules for arbitrary $\Sigma$, without requiring knowledge of the exact eigenvalues, which is not possible in general, and without including the zero mode.

As discussed in Ref.~\cite{Amore10}, one can define $\Phi_n=\sqrt{\Sigma}\,\psi_n$ and cast Eq.~(\ref{eq_Helmoltz_dens}) into the equivalent form
\begin{equation}
\frac{1}{\sqrt{\Sigma}} (-\Delta) \frac{1}{\sqrt{\Sigma}} \Phi_n(\theta,\phi) = E_n \Phi_n(\theta,\phi)
\label{eq_Helmoltz_dens2}
\end{equation} 
in terms of the Hermitian operator $\hat{O} \equiv \frac{1}{\sqrt{\Sigma}} (-\Delta) \frac{1}{\sqrt{\Sigma}}$. 

Since the lowest eigenvalue of $\hat{O}$ vanishes, it is convenient to introduce the modified operator, following Ref.~\cite{Amore14},
\begin{equation}
\hat{O}_\gamma \equiv \frac{1}{\sqrt{\Sigma}} (-\Delta +\gamma) \frac{1}{\sqrt{\Sigma}} \ ,
\end{equation}
where $\gamma$ is a constant parameter that will eventually be sent to zero.

The Green's function associated with $-\Delta + \gamma$ is 
\begin{equation}
\begin{aligned}
G^{(d)}_\gamma (\Omega,\Omega')  &= -\mathcal{G}^{(d)}(-\gamma, {\bf n}, {\bf n'})  \\
&= \frac{1}{\gamma \mathrm{Vol}(S^{d})}  + 
\sum_{\ell=1}^{\infty}\frac{1}{\ell(\ell+d-1) + \gamma}  \sum_{\vec{m}} Y_{\ell, \vec{m}}(\Omega)\, Y^*_{\ell,\vec{m}}(\Omega') ,
\end{aligned}
\label{eq_green_Gd}
\end{equation}
and it contains the divergent contribution of the zero mode in the limit $\gamma \rightarrow 0^+$.

Expanding for $\gamma \rightarrow 0^+$, we obtain
\begin{equation}
\begin{aligned}
G^{(d)}_\gamma (\Omega,\Omega') &= \frac{1}{\gamma \mathrm{Vol}(S^{d})}  + \sum_{p=0}^\infty (-\gamma)^p 
\sum_{\ell=1}^{\infty} \left( \frac{1}{\ell(\ell+d-1)}\right)^{p+1}  \sum_{\vec{m}} Y_{\ell, \vec{m}}(\Omega)\, Y^*_{\ell,\vec{m}}(\Omega') \\
&= \frac{1}{\gamma \mathrm{Vol}(S^{d})}  + \sum_{p=0}^\infty (-\gamma)^p G^{(d,p)}(\Omega,\Omega') \ .
\end{aligned}
\label{eq_green_Gd_2}
\end{equation}

The Green's function associated with $\hat{O}_\gamma$ can then be expressed as
\[
G^{(d)}_{\hat{O}_\gamma}(\Omega,\Omega')  = \sqrt{\Sigma(\Omega)}  \ G^{(d)}_\gamma (\Omega,\Omega')   \ \sqrt{\Sigma(\Omega')} \ .
\]

We then define the sum rule for the eigenvalues of $\hat{O}_\gamma$:
\[
Z_p^{(d)}(\gamma) \equiv \sum_{n=0}^\infty \frac{1}{\left( E_n^{(d)(\gamma)}\right)^p} \ ,
\]
which, for $\gamma>0$, can be expressed as the trace
\[
Z_p^{(d)}(\gamma) = \int G^{(d)}_{\hat{O}_\gamma}(\Omega_1,\Omega_2) \dots  G^{(d)}_{\hat{O}_\gamma}(\Omega_{p},\Omega_1)  \ d\Omega_1 \dots d\Omega_p \ . 
\]

Notice that $Z_p^{(d)}(\gamma)$ diverges as $\gamma \rightarrow 0^+$ because of the contribution of the zero mode; for this reason we introduce the \emph{renormalized} sum rule
\[
\tilde{Z}_p^{(d)}(\gamma) \equiv \sum_{n=1}^\infty \frac{1}{\left( E_n^{(d)}(\gamma)\right)^p}  = Z_p^{(d)}(\gamma) - \frac{1}{\left( E_0^{(d)}(\gamma)\right)^p} 
\]
and proceed to verify that $\tilde{Z}_p^{(d)}(\gamma)$ remains finite in the limit $\gamma \rightarrow 0^+$. 

From Eq.~(\ref{eq_green_Gd}) it is clear that the trace $Z_p^{(d)}(\gamma)$ contains $p$ terms that diverge as $\gamma \rightarrow 0^+$, together with a finite term:
\[
\begin{aligned}
Z_p^{(d)}(\gamma) 
&=  \frac{1}{\gamma^p} \left[\frac{\int \Sigma(\Omega) d\Omega }{\mathrm{Vol}(S^{d})} \right]^p \\
&+  \frac{p}{\gamma^{p-1}}  \frac{\left( \int \Sigma(\Omega) d\Omega \right)^{p-2}}{\left( \mathrm{Vol}(S^{d}) \right)^{p-1}} \  \int \Sigma(\Omega) G^{(d)}(\Omega, \Omega') \Sigma(\Omega') d\Omega d\Omega' + \dots \\
\end{aligned}
\]

In particular, one can work out the explicit expressions of $Z_p^{(d)}$ for $p=2$ and $p=3$:
\begin{equation}
\begin{aligned}
Z_2^{(d)}(\gamma)  &= \frac{1}{\gamma^2} \left( \frac{\int \Sigma(\Omega) d\Omega}{\mathrm{Vol}(S^{d})} \right)^2 + 
\frac{2}{\gamma} \frac{\int \Sigma(\Omega) G^{(d,0)}(\Omega,\Omega') \Sigma(\Omega') d\Omega d\Omega'}{\mathrm{Vol}(S^{d})} \\
&+ \int \Sigma(\Omega) G^{(d,0)}(\Omega,\Omega') \Sigma(\Omega') G^{(d,0)}(\Omega,\Omega') d\Omega d\Omega' \\
&- \frac{2}{\mathrm{Vol}(S^{d})} \int \Sigma(\Omega) G^{(d,1)}(\Omega,\Omega') \Sigma(\Omega') d\Omega d\Omega' + \dots\\
Z_3^{(d)}(\gamma)  &= \frac{1}{\gamma^3} \left( \frac{\int \Sigma(\Omega) d\Omega}{\mathrm{Vol}(S^{d})} \right)^3  + 
\frac{3}{\gamma^2} \frac{\int \Sigma(\Omega) d\Omega }{\mathrm{Vol}(S^{d})^2}  \ \int \Sigma(\Omega) G^{(d,0)}(\Omega,\Omega') \Sigma(\Omega') d\Omega d\Omega' \\
&+  \frac{3}{\gamma} \left[  -\frac{\int \Sigma(\Omega) d\Omega}{\mathrm{Vol}(S^{d})^2}   
\int \Sigma(\Omega) G^{(d,1)}(\Omega,\Omega') \Sigma(\Omega') d\Omega d\Omega'
\right. \\
&+ \left. \frac{1}{\mathrm{Vol}(S^{d})}   
\int \Sigma(\Omega) G^{(d,0)}(\Omega,\Omega') \Sigma(\Omega') G^{(d,0)}(\Omega',\Omega'') \Sigma(\Omega'') d\Omega d\Omega' d\Omega''
\right] \\
&+ \int \Sigma(\Omega) G^{(d,0)}(\Omega,\Omega') \Sigma(\Omega') G^{(d,0)}(\Omega',\Omega'') \Sigma(\Omega'')  G^{(d,0)}(\Omega'',\Omega''')d\Omega d\Omega' d\Omega'' d\Omega''' \\
&+ \frac{3}{\mathrm{Vol}(S^{d})^2} \int \Sigma(\Omega) d\Omega   \ \int \Sigma(\Omega) G^{(d,2)}(\Omega,\Omega') \Sigma(\Omega') d\Omega d\Omega' \\
&- \frac{6}{\mathrm{Vol}(S^{d})} \int \Sigma(\Omega) G^{(d,1)}(\Omega,\Omega') \Sigma(\Omega')  G^{(d,0)}(\Omega',\Omega'') \Sigma(\Omega'') d\Omega d\Omega'  + \dots\\
\end{aligned}
\end{equation}

On the other hand, the energy of the lowest mode behaves as
\[
E_0(\gamma)  =  \gamma \epsilon_1 + \gamma^2 \epsilon_2 + \dots \ ,
\]
where the corrections $\epsilon_1$, $\epsilon_2$, \dots\ have been calculated explicitly in Appendix~\ref{app:perturbation_theory} using perturbation theory up to order $4$.

For arbitrary integer values of $p$ we have 
\[
\begin{aligned}
\frac{1}{E_0(\gamma)^p} &= \frac{1}{\gamma ^{p}}  \frac{1}{\epsilon _1^{p}}  -\frac{p}{\gamma ^{p-1}} \frac{\epsilon_2}{\epsilon _1^{p+1}}
+\frac{1}{2} \frac{p}{\gamma^{p-2}}    \frac{\left((1+p) \epsilon _2^2-2 \epsilon _1 \epsilon_3\right)}{\epsilon _1^{p+2}}  \\
 &-\frac{1}{6} \frac{p}{\gamma ^{p-3}}  \frac{\left((1+p) (2+p) \epsilon _2^3-6 (1+p) \epsilon _1 \epsilon _2 \epsilon _3+6 \epsilon _1^2 \epsilon _4\right)}{\epsilon _1^{p+3}} +\dots \ .
\end{aligned}
\]

For $p=2,3$ this expression reduces to
\[
\begin{aligned}
\left[ \frac{1}{ E_0(\gamma)} \right]^2 &= \frac{1}{\gamma ^2 \epsilon _1^2}-\frac{2 \epsilon _2}{\gamma  \epsilon _1^3}+\frac{3 \epsilon _2^2-2 \epsilon _1 \epsilon _3}{\epsilon _1^4} + O(\gamma) \\
\left[ \frac{1}{ E_0(\gamma)} \right]^3 &= \frac{1}{\gamma ^3 \epsilon _1^3}-\frac{3 \epsilon _2}{\gamma ^2 \epsilon _1^4}
-\frac{3 \left(\epsilon _1 \epsilon _3-2 \epsilon_2^2\right)}{\gamma  \epsilon _1^5}
-\frac{10 \epsilon _2^3-12 \epsilon _1 \epsilon _3 \epsilon _2+3 \epsilon _1^2 \epsilon_4}{\epsilon _1^6} + O(\gamma) \\
\end{aligned} \ .
\]

By substituting the explicit expressions for $\epsilon_1,\dots,\epsilon_4$, derived in Appendix~\ref{app:perturbation_theory}, one verifies that every divergent contribution in $\left[E_0(\gamma)^{-1}\right]^p$ for $p=2,3$ is matched by a corresponding divergent term in the trace. The divergences therefore cancel completely, and the sum rules take the form:
\begin{equation}
\begin{aligned}
\sum_{n=1}^\infty \frac{1}{E_n^2} &= \lim_{\gamma \rightarrow 0^+} \tilde{Z}_2(\gamma) \\
&= \int \Sigma(\Omega) G^{(d,0)}(\Omega,\Omega') \Sigma(\Omega') G^{(d,0)}(\Omega',\Omega) d\Omega d\Omega' \\
&- \frac{2}{\mathrm{Vol}(S^{d})} \int \Sigma(\Omega) G^{(d,1)}(\Omega,\Omega') \Sigma(\Omega') d\Omega d\Omega' - \frac{3 \epsilon_2^2 -2\epsilon_1 \epsilon_3}{\epsilon_1^4} \\
\sum_{n=1}^\infty \frac{1}{E_n^3} &= \lim_{\gamma \rightarrow 0^+} \tilde{Z}_3(\gamma) \\
&= \int \Sigma(\Omega) G^{(d,0)}(\Omega,\Omega') \Sigma(\Omega') G^{(d,0)}(\Omega',\Omega'') \Sigma(\Omega'')  G^{(d,0)}(\Omega'',\Omega''')d\Omega d\Omega' d\Omega'' d\Omega''' \\
&+ \frac{3}{\mathrm{Vol}(S^{d})^2} \int \Sigma(\Omega) d\Omega   \ \int \Sigma(\Omega) G^{(d,2)}(\Omega,\Omega') \Sigma(\Omega') d\Omega d\Omega' \\
&- \frac{6}{\mathrm{Vol}(S^{d})} \int \Sigma(\Omega) G^{(d,1)}(\Omega,\Omega') \Sigma(\Omega')  G^{(d,0)}(\Omega',\Omega'') \Sigma(\Omega'') d\Omega d\Omega'   \\
&+ \frac{10 \epsilon _2^3-12 \epsilon _1 \epsilon _3 \epsilon _2+3 \epsilon _1^2 \epsilon_4}{\epsilon _1^6} 
\end{aligned}
\label{eq:sum_rules_2_3}
\end{equation}

The same cancellation mechanism must hold for arbitrary integers $p\geq p_{\rm min}$, however checking this explicitly requires working in perturbation theory through order $p+1$.  

Without loss of generality we consider a density of the general form
\begin{equation}
\Sigma(\Omega) = 1 + \sum_{l=1}^\infty \sum_{\vec{m}}^l c_{l\vec{m}} Y_{l\vec{m}}(\Omega) \ ,
\end{equation}
where $c_{l\vec{m}}$ are arbitrary coefficients such that $\Sigma(\Omega) > 0$ over the $d$-sphere.

In this case the total mass is simply given by
\begin{equation}
\int \Sigma(\Omega) d\Omega = \mathrm{Vol}(S^{d}) \ .
\end{equation}

We define the integrals~\footnote{To avoid heavy notation, we omit the superscript $d$ in the symbol representing the integral.}:
\begin{subequations}
\begin{align}
\mathcal{I}_1^{(q)} &\equiv \int \Sigma(\Omega) G^{(d,q)}(\Omega,\Omega') \Sigma(\Omega')  d\Omega d\Omega' \nonumber \\
\mathcal{I}_2^{(q,p)} &\equiv \int \Sigma(\Omega) G^{(d,q)}(\Omega,\Omega') \Sigma(\Omega')  G^{(d,p)}(\Omega',\Omega'') \Sigma(\Omega'') d\Omega d\Omega' d\Omega'' \nonumber \\
\mathcal{I}_3^{(q,p,r)} &\equiv \int \Sigma(\Omega) G^{(d,q)}(\Omega,\Omega') \Sigma(\Omega')  G^{(d,p)}(\Omega',\Omega'') \Sigma(\Omega'') G^{(d,r)}(\Omega'',\Omega''') \Sigma(\Omega''') d\Omega d\Omega' d\Omega''  d\Omega'''\nonumber \\
\mathcal{J}_1^{(q, p)} &\equiv \int \Sigma(\Omega) G^{(d,q)}(\Omega,\Omega') \Sigma(\Omega') G^{(d,p)}(\Omega',\Omega) d\Omega d\Omega' \nonumber  \\
\mathcal{J}_2^{(q, p,r)} &\equiv \int \Sigma(\Omega) G^{(d,q)}(\Omega,\Omega') \Sigma(\Omega') G^{(d,p)}(\Omega',\Omega'') \Sigma(\Omega'') G^{(d,r)}(\Omega'',\Omega) d\Omega d\Omega' d\Omega'' \nonumber  
\end{align}
\end{subequations}

The explicit expressions of these integrals in terms of the expansion coefficients $c_{l\vec{m}}$ are reported in Appendix~\ref{appB}. We can cast the sum rules (\ref{eq:sum_rules_2_3}) in terms of these integrals as
\begin{equation}
\begin{aligned}
\sum_{n=1}^\infty \frac{1}{E_n^2} &= \mathcal{J}_1^{(0,0)} +  \left( \frac{\mathcal{I}_1^{(0)}}{\mathrm{Vol}(S^{d}) } \right)^2 - 2 \frac{\mathcal{I}_2^{(0,0)}}{\mathrm{Vol}(S^{d}) } \\
\sum_{n=1}^\infty \frac{1}{E_n^3} &=  \mathcal{J}_2^{(0,0,0)} - \left( \frac{\mathcal{I}_1^{(0)}}{\mathrm{Vol}(S^{d}) } \right)^3 
+ 3 \frac{\mathcal{I}_1^{(0)} \ \mathcal{I}_2^{(0,0)}}{\left(\mathrm{Vol}(S^{d})\right)^2}
- 3\frac{\mathcal{I}_3^{(0,0,0)}}{\mathrm{Vol}(S^{d}) } \\
\end{aligned}
\label{eq_zeta_2_3}
\end{equation}

Equations (\ref{eq_zeta_2_3}) express the spectral sum rules as integral functionals of the density and of the Green's functions 
$G^{(d,q)}(\Omega,\Omega')$ and can be cast as algebraic expressions in terms of the coefficients $c_{\ell, \textbf{m}}$ of the spectral decomposition
of the density (see the Appendix). A simple case which can be calculated exactly is discussed in the next Section.

\section{An application}
\label{sec:application}

We now consider an illustrative  application of the formulas derived in this paper, corresponding to the density
\begin{equation}
\Sigma(\Omega) = 1 + \kappa Y_{1,\textbf{0}}(\Omega) \ ,
\end{equation}
which, for $d=2$, reduces to the case studied in Ref.~\cite{Amore20}.  The condition $\Sigma>0$ requires $|\kappa|<\sqrt{\mathrm{Vol}(S^d)/(d+1)}$.

\begin{table}[h]
\centering
\setlength{\tabcolsep}{4pt}
\renewcommand{\arraystretch}{1.2}
\resizebox{\textwidth}{!}{%
\begin{tabular}{|c|c|c|c|c|}
\hline
 & $d=2$ & $d=3$ & $d=4$ & $d=5$ \\
\hline
$\mathcal{I}_1^{(0)}$
& $\frac{\kappa^2}{2}$
& $\frac{\kappa^2}{3}$
& $\frac{\kappa^2}{4}$
& $\frac{\kappa^2}{5}$ \\
\hline
$\mathcal{I}_2^{(0,0)}$
& $\frac{\kappa^2}{4}$
& $\frac{\kappa^2}{9}$
& $\frac{\kappa^2}{16}$
& $\frac{\kappa^2}{25}$ \\
\hline
$\mathcal{I}_3^{(0,0,0)}$
& $\frac{\kappa^2}{8} + \frac{\kappa^4}{120\pi}$
& $\frac{\kappa^2}{27} + \frac{\kappa^4}{144\pi^2}$
& $\frac{\kappa^2}{64} + \frac{3\kappa^4}{1120\pi^2}$
& $\frac{\kappa^4}{240\pi^3} + \frac{\kappa^2}{125}$ \\
\hline
$\mathcal{J}_1^{(0,0)}$
& $1+\frac{\kappa^2}{8\pi}$
& $\frac{3+4\pi^2}{48}+\frac{11\kappa^2}{36\pi^2}$
& $-$
& $-$ \\
\hline
$\mathcal{J}_2^{(0,0,0)}$
& $2(\zeta(3)-1)+\frac{3\kappa^2}{32\pi}$
& $\frac{2\pi^2-3}{96}+\frac{(4\pi^2-29)\kappa^2}{96\pi^2}$
& $\frac{277\kappa^2}{4608\pi^2}+\frac{2\zeta(3)}{27}+\frac{23}{1458}$
& $\left(\frac{2833}{96000\pi^3}+\frac{1}{80\pi}\right)\kappa^2+\frac{15+152\pi^2}{18432}$ \\
\hline
\end{tabular}%
}
\caption{Values of the integrals entering the sum rules.}
\label{tab:table_integrals}
\end{table}

Using the results reported in Table~\ref{tab:table_integrals}, we can write the sum rules of order two and three explicitly for $d=3,4,5$ dimensions. Notice, however, that the sum rule of order two is finite only for $d=3$.

For $d=3$ we have:
\begin{equation}
\begin{aligned}
\sum \frac{1}{E_n^2} &= \frac{3+4 \pi ^2}{48} +\frac{7 \kappa ^2}{36 \pi^2}+\frac{\kappa ^4}{36 \pi ^4} \\
\sum \frac{1}{E_n^3} &= \frac{-3+2 \pi ^2}{96} +\left(\frac{1}{24}-\frac{103}{288
   \pi ^2}\right) \kappa ^2+\frac{5 \kappa ^4}{288 \pi ^4}-\frac{\kappa
   ^6}{216 \pi ^6}
\end{aligned}
\end{equation}

For $d=4$ we have:
\begin{equation}
\sum \frac{1}{E_n^3} = \frac{23}{1458}+\frac{2 \zeta (3)}{27}+\frac{49 \kappa ^2}{1152 \pi ^2}+\frac{513
   \kappa ^4}{143360 \pi ^4}-\frac{27 \kappa ^6}{32768 \pi
   ^6}
\end{equation}

For $d=5$ we have:
\begin{equation}
\sum \frac{1}{E_n^3} = \frac{15+152 \pi ^2}{18432}
+\left(\frac{529}{96000 \pi ^3}+\frac{1}{80 \pi }\right) \kappa^2
+\frac{23 \kappa ^4}{2000 \pi^6}
-\frac{\kappa ^6}{125 \pi ^9}
\end{equation}

To verify these formulas, we follow the same approach used in Ref.~\cite{Amore20} for the case $d=2$.

We apply the Rayleigh--Ritz method with a truncation at $\ell = \ell_{\rm max}$, working in a subspace of the Hilbert space with 
\[
\overline{N}^{(d)}(\ell_{\rm max}) \equiv \sum_{\ell=0}^{\ell_{\rm max}}  g_\ell^{(d)} = \frac{\left(d+2 \ell _{\max }\right) \Gamma \left(d+\ell _{\max}\right)}{\Gamma (d+1) \Gamma \left(\ell _{\max }+1\right)}
\]
elements.
Since $\overline{N}^{(d)}(\ell_{\rm max})  \approx \frac{2}{d!} \ell_{\rm max}^d$ for $\ell_{\max} \gg 1$, reaching large values of $\ell_{\rm max}$ becomes increasingly difficult as $d$ grows. The largest value used in Ref.~\cite{Amore20}, namely $\ell_{\rm max}=90$, required a subspace with $8281$ elements for $d=2$. In $d=3$, $d=4$, and $d=5$, however, the number of elements increases to $255346$, $5969236$, and $112831537$, respectively. 

Because matrices of such size are not tractable in practice, we restrict ourselves to $\ell_{\rm max} = 30$ for $d=3$ ($10416$ elements), $\ell_{\rm max} = 20$ for $d=4$ ($19481$ elements), and $\ell_{\rm max} = 15$ for $d=5$ ($27132$ elements). In practice, one should not use the full set of numerical eigenvalues produced by the calculation, since the truncation affects the higher part of the spectrum more severely. A reasonable compromise is therefore to retain only the lowest half of the computed eigenvalues.

The remaining part of the spectrum can then be approximated using Weyl's law, which becomes increasingly accurate for $\ell \gg \ell_{\rm max}$. Theorem 1.1 of Ref.~\cite{Bandara21} implies that the number of eigenvalues below a given value $\Lambda$ behaves as
\begin{equation}
\label{eq:weyl-law-standard}
N(\Lambda)
\sim
\frac{1}{(4\pi)^{d/2}\Gamma\!\left(1+\frac d2\right)}
\left(
\int_{S^d}\Sigma(\Omega)^{d/2}\,d\Omega_d
\right)
\Lambda^{d/2}.
\end{equation}

In particular, for $d=3,4,5$ we obtain
\begin{equation}
\begin{aligned}
E_n^{(3)} &\approx
\frac{ 2^{1/6}\,3^{2/3}\,\pi\, n^{2/3}}{\bigl(4\kappa^3  + 6\sqrt{2}\,\pi\,\kappa^2+ 6\pi^2\kappa    + \sqrt{2}\,\pi^3\bigr)^{1/3}\,F_3(\kappa)^{2/3}},   \\[1.2ex]
E_n^{(4)} &\approx 4\sqrt{6}\,\pi\, \sqrt{\frac{n}{3\kappa^2+8\pi^2}} \\[1.2ex]
E_n^{(5)} &\approx \frac{60^{2/5}\,\pi^{3/2}\,n^{2/5}}{\bigl(\pi^{3/2}+\sqrt{6}\,\kappa\bigr)\, F_5(\kappa)^{2/5}}.
\end{aligned}
\end{equation}
where 
\begin{equation}
\begin{split}
F_3(\kappa) &\equiv {}_2F_1\!\left( -\frac{3}{2}, \frac{3}{2}; 3; \frac{2\sqrt{2}\,\kappa}{\sqrt{2}\,\kappa+\pi} \right) \\
F_5(\kappa) &\equiv {}_2F_1\!\left(-\frac{5}{2}, \frac{5}{2}; 5; \frac{2\sqrt{6}\,\kappa}{\pi^{3/2}+\sqrt{6}\,\kappa} \right) \\
\end{split}
\end{equation}

\begin{figure}[t]
    \centering
    \includegraphics[width=0.6\textwidth]{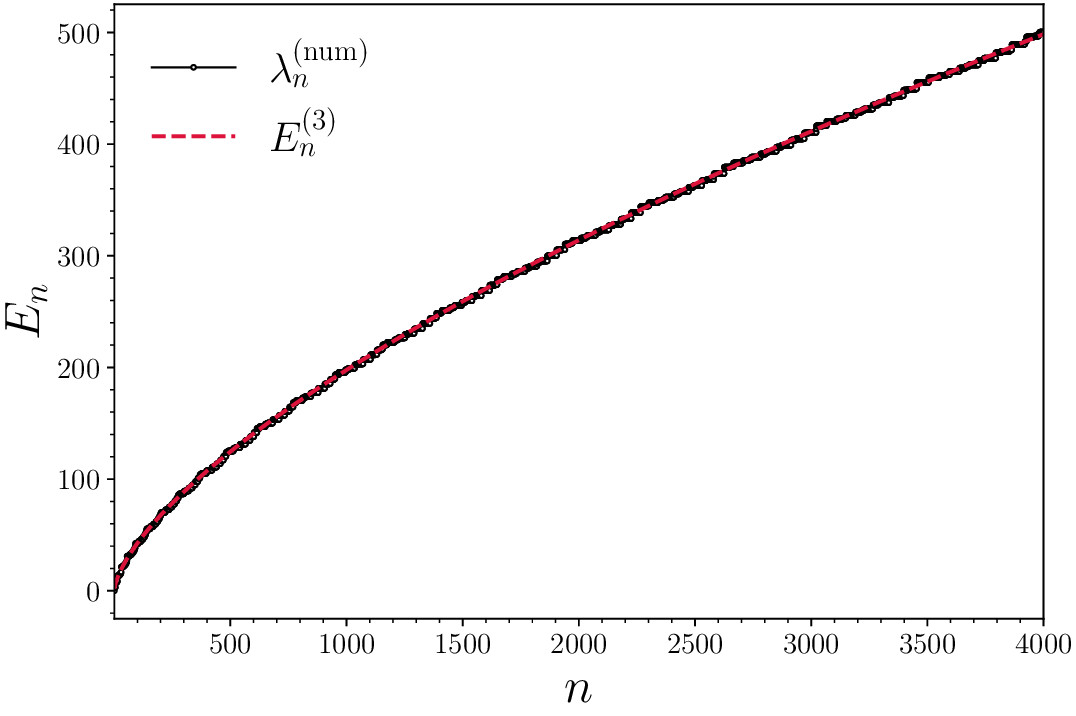}
    \caption{Comparison between the numerical eigenvalues on the $3$--sphere with $\ell_{\rm max}=30$ and $\kappa=2$ and the Weyl-law approximation.}
    \label{fig:eigenvalues_weyl_d3}
\end{figure}

As an example, in Fig.~\ref{fig:eigenvalues_weyl_d3} we plot the numerical eigenvalues on the $3$-sphere obtained with the Rayleigh--Ritz method with $\ell_{\rm max}=30$ 
and compare them with the behavior obtained from Weyl's law. 

\begin{figure}[t]
    \centering
    \includegraphics[width=0.6\textwidth]{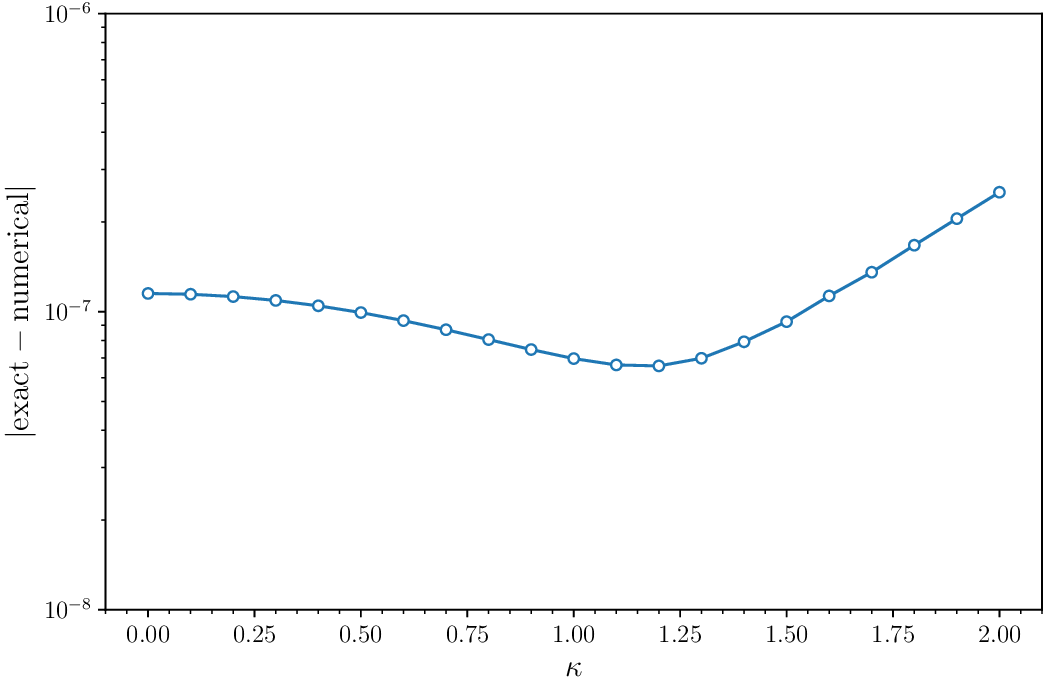}
    \caption{Difference between the exact sum rule for $d=3$ and $p=3$ and the numerical approximation obtained by using the Rayleigh--Ritz method with $\ell_{\rm max}=30$ and Weyl's law.}
    \label{fig:exact_vs_numerical_d3}
\end{figure}

In Fig.~\ref{fig:exact_vs_numerical_d3} we then plot the difference  between the exact sum rule for $d=3$ and $p=3$ and the numerical approximation obtained by using the Rayleigh--Ritz method with $\ell_{\rm max}=30$ and Weyl's law.

\begin{figure}[t]
    \centering
    \includegraphics[width=0.6\textwidth]{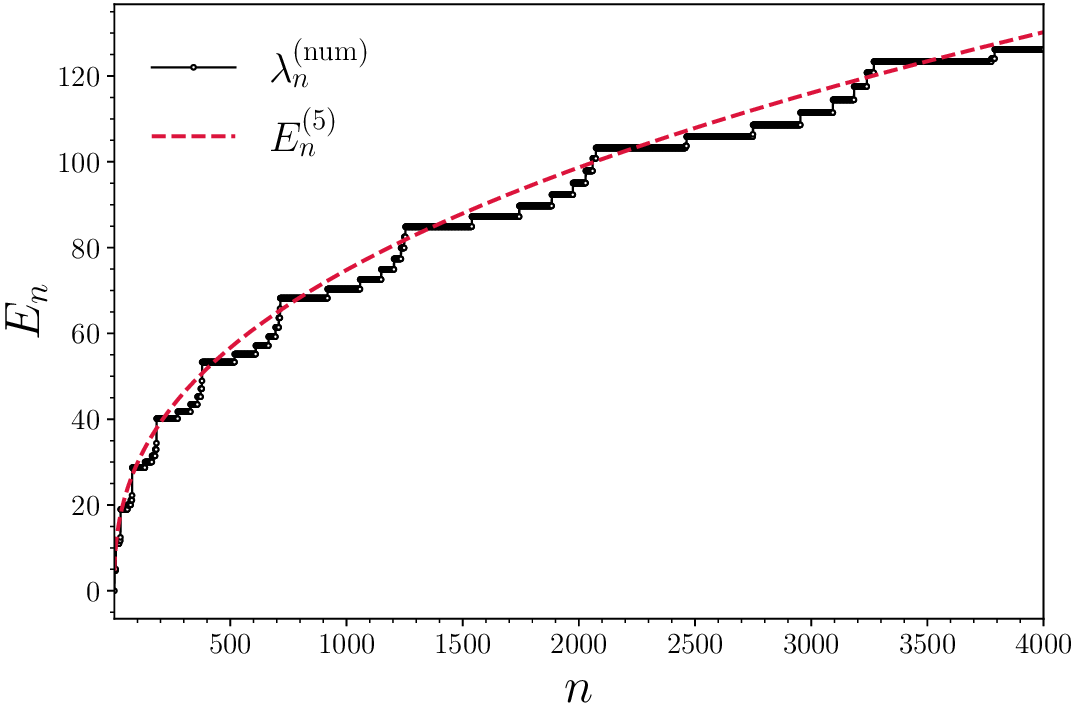}
    \caption{Comparison between the numerical eigenvalues on the $5$--sphere with $\ell_{\rm max}=15$ and $\kappa=2$ and the Weyl-law approximation.}
    \label{fig:eigenvalues_weyl_d5}
\end{figure}

As commented earlier, at fixed cutoff $\ell_{\rm max}$ there is an exponential growth in the size of the matrices with the dimension of the sphere, which forces one to use smaller cutoffs when increasing the dimensions. Because of this, at larger $d$  the value of the cutoff may not be sufficiently large for the asymptotic regime described by Weyl's law to apply: this is illustrated in Fig.~\ref{fig:eigenvalues_weyl_d5} for the case of the $5$--sphere, using $\ell_{\rm max}=15$ and $\kappa=2$.

As a result, the difference between the exact sum rule of order $3$ and the corresponding numerical approximation is larger than in the previous case.

\begin{figure}[t]
    \centering
    \includegraphics[width=0.6\textwidth]{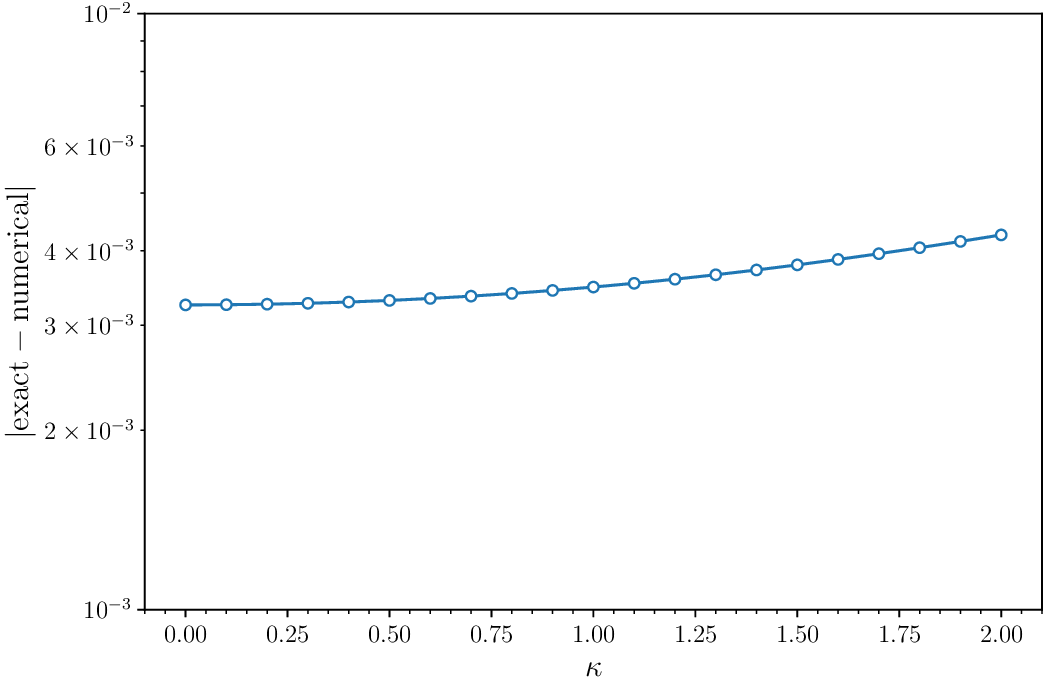}
    \caption{Difference between the exact sum rule for $d=5$ and $p=3$ and the numerical approximation obtained by using the Rayleigh--Ritz method with $\ell_{\rm max}=15$ and Weyl's law.}
    \label{fig:exact_vs_numerical_d5}
\end{figure}

Actually, obtaining an estimate of the spectral sum rule for large dimensions is even more challenging than what we just stated: holding $\ell_{\rm max}$ fixed while increasing the dimensions of the sphere leads to an exponential growth in matrix size, but  it is not sufficient to retain the same level of accuracy, because the asymptotic regime
is pushed farther away. To see this, we can focus on the case $\kappa=0$, for which the eigenvalues are known explicitly: as we can see from Figs.~\ref{fig:exact_vs_numerical_d3} and \ref{fig:exact_vs_numerical_d5}, the difference between the exact sum rule and the numerical estimate varies moderately with $\kappa$ in the region of values considered. 
This occurs because the main source of error for $\kappa >0$ is not the numerical estimate of the eigenvalues but the approximation of the tail of the series using Weyl's law.

\begin{figure}[t]
 \centering
 \includegraphics[width=0.6\textwidth]{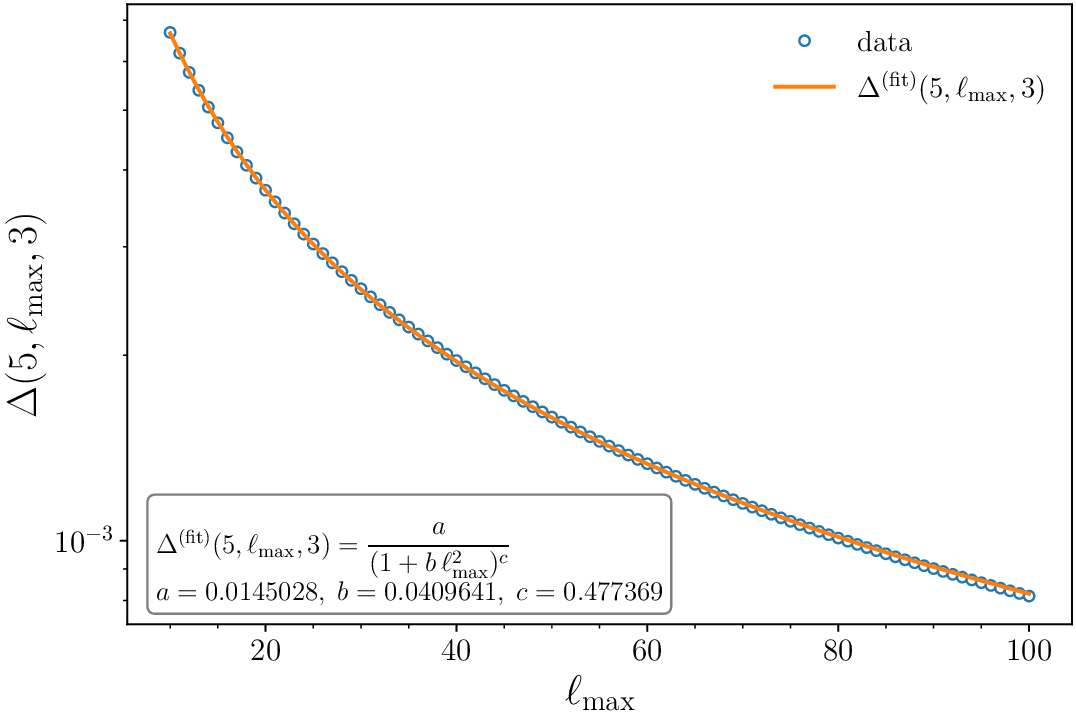}
\caption{Behavior of $\Delta(5,\ell_{\rm max},3)$ as a function of $\ell_{\rm max}$, where $\Delta(d,\ell_{\rm max},s)$ measures the difference between the exact tail contribution and its Weyl-law approximation (here for $\kappa=0$).}
 \label{fig:delta}
\end{figure}

 In Fig.~\ref{fig:delta}  we plot  
 \[
 \Delta(d,\ell_{\rm max},s) \equiv \left| \sum_{\ell=\ell_{\rm max}+1}^\infty \frac{g_\ell^{(d)}}{(\ell (\ell+d-1))^s}  - \sum_{n=\overline{N}^{(d)}(\ell_{\rm max})}^\infty \frac{1}{E_n^s}\right|
 \ . \] 
 Clearly $\lim_{\ell_{\rm max}\rightarrow\infty} \Delta(d,\ell_{\rm max},s) = 0$. 

As we can see from the plot, the fit
\[
 \Delta^{\rm (fit)}(5,\ell_{\rm max},3) \approx \frac{0.0145028}{\left(1+0.0409641 \ell_{\rm max}^2\right)^{0.477369}}
\]
describes very well the behavior over the range of values considered and it can be used to estimate (roughly) the value of $\ell_{\rm max}$ needed to 
achieve a given precision. Remarkably, we see that  $\Delta^{\rm (fit)}(5,\ell_{\rm max},3) \approx 10^{-6}$ for $\ell_{\rm max} = 10^5$. In this case however
$\overline{N}^{(d)}(10^5) \approx 10^{23}$, which is clearly completely out of reach.

\section{Conclusions}
\label{conclusions}

We have developed a rigorous exact framework to calculate spectral sum rules associated with
inverse powers of the eigenvalues of the weighted Laplacian on a $d$-sphere in the presence
of an arbitrary positive density. The method is based on a renormalized trace formulation:
the sum rules are expressed as traces in the basis of hyperspherical harmonics, while the
divergent contribution of the zero mode is removed through a systematic subtraction based on
its perturbative expansion in an auxiliary shift parameter. In this way, one obtains exact
expressions for the renormalized sum rules without requiring explicit knowledge of the
spectrum.

We have worked out this construction explicitly for the renormalized sum rules of orders  $2$ and $3$, and
we have illustrated the formalism in closed form for the density
$\Sigma(\Omega)=1+\kappa Y_{1,\vec 0}(\Omega)$ in $d=3,4,5$. In particular, the example of
Section \ref{sec:application} is meant to illustrate the implementation of the general framework.

We have also provided a qualitative numerical consistency check of the formalism by comparing
the exact expressions with estimates obtained from a Rayleigh--Ritz/Weyl approach.
As the dimension increases, however, the agreement deteriorates because this numerical
strategy becomes less efficient: for fixed cutoff $\ell_{\max}$, the Rayleigh--Ritz subspace is
not sufficiently large to approach the asymptotic regime, where Weyl's law provides an
accurate description of the spectral tail.  The numerical approach therefore suffers from a form of dimensional curse,
which becomes increasingly severe as $d$ grows.

Several directions for future work appear naturally:
\begin{itemize}
\item deriving explicit expressions for sum rules of higher order, which would require extending
the perturbative treatment of the zero mode to correspondingly higher orders;

\item using the exact sum rules to probe the asymptotic behavior of the spectrum beyond the leading contribution (Weyl's law);

\item extending the method to weighted eigenvalue problems on more general manifolds,
provided that a suitable orthonormal and complete basis is available;
\end{itemize}

\section*{Acknowledgements}
The research of P.A. was supported by the Sistema Nacional de Investigadores (M\'exico).

ChatGPT was used solely for language editing and as a programming aid. All scientific content, derivations, numerical analyses and conclusions are the sole responsibility of the author.

\begin{appendix}
\numberwithin{equation}{section}

\section{Green's functions}
We report here the expressions for $G^{(d)}_\gamma(\Omega,\Omega')$ about $\gamma=0$, up to order $\gamma^2$: the higher order Green's functions for $d=2$,$3$ and $4$
can be read off by isolating the coefficients of powers of $\gamma$ (up to order $\gamma^2$):
\begin{equation}
\begin{aligned}
G^{(2)}_\gamma(\Omega,\Omega')
&= \frac{1}{4 \pi \gamma}
 - \frac{\log \left(\sin^2\left(\frac{\theta}{2}\right)\right)+1}{4 \pi} \\
&\quad
+ \frac{\gamma}{24 \pi}
\left(
-6 \text{Li}_2\left(\cos^2\left(\frac{\theta}{2}\right)\right)
+ \pi^2 - 6
\right) \\
&\quad
+ \frac{\gamma^2}{24 \pi}
\Biggl(
-12 \text{Li}_3\left(\sin^2\left(\frac{\theta}{2}\right)\right)
- 6 \text{Li}_2\left(\cos^2\left(\frac{\theta}{2}\right)\right) \\
&\qquad\qquad
+ 6 \log\left(\sin^2\left(\frac{\theta}{2}\right)\right)
  \text{Li}_2\left(\sin^2\left(\frac{\theta}{2}\right)\right)
+ 12 \zeta(3) + \pi^2 - 12
\Biggr)
+ O(\gamma^3) \\
\\
G^{(3)}_\gamma(\Omega,\Omega')
&= \frac{1}{2 \pi^2 \gamma}
+ \frac{2(\pi-\theta)\cot(\theta)-1}{8 \pi^2} \\
&\quad
- \frac{\gamma}{96 \pi^2}
\left(
6 \theta^2 - 12 \pi \theta + 4 \pi^2 + 3
\right) \\
&\quad
+ \frac{\gamma^2}{192 \pi^2}
\Biggl(
-3\left(\theta^2+1\right)
+ 6 \pi \theta
+ 2(\theta-2\pi)(\theta-\pi)\theta \cot(\theta)
- 2 \pi^2
\Biggr) + O(\gamma^3) \\
\\
G^{(4)}_\gamma(\Omega,\Omega')
&= \frac{3}{8 \pi^2 \gamma}
+ \frac{-7 \cos(\theta) - 6(\cos(\theta)-1)\log\left(\sin\left(\frac{\theta}{2}\right)\right) + 4}
       {24 \pi^2 (\cos(\theta)-1)} \\
&\quad
+ \frac{\gamma}{432 \pi^2 (\cos(\theta)+1)}
\Biggl(
\left(3 \pi^2 - 2\right)(\cos(\theta)+1)
+ 36(2\cos(\theta)+1)\log\left(\sin\left(\frac{\theta}{2}\right)\right) \\
&\qquad\qquad
- 18(\cos(\theta)+1)\text{Li}_2\left(\cos^2\left(\frac{\theta}{2}\right)\right)
\Biggr) \\
&\quad
+ \frac{\gamma^2}{7776 \pi^2}
\Biggl(
9 \pi^2 \csc^2\left(\frac{\theta}{2}\right)
+ 144 \log\left(\sin\left(\frac{\theta}{2}\right)\right)
+ \frac{72 \log\left(\csc\left(\frac{\theta}{2}\right)\right)}{\cos(\theta)+1} \\
&\qquad\qquad
- 216 \text{Li}_3\left(\sin^2\left(\frac{\theta}{2}\right)\right)
+ 36\left(\frac{3}{\cos(\theta)-1}+5\right)
  \text{Li}_2\left(\cos^2\left(\frac{\theta}{2}\right)\right) \\
&\qquad\qquad
+ 216 \log\left(\sin\left(\frac{\theta}{2}\right)\right)
  \text{Li}_2\left(\sin^2\left(\frac{\theta}{2}\right)\right)
+ 216 \zeta(3) - 30 \pi^2 - 8
\Biggr) + O(\gamma^3) .
\end{aligned}
\end{equation}

\section{Perturbation theory for the zero mode}
\label{app:perturbation_theory}

The eigenvalue equation for the zero mode is
\[
(-\Delta_{S^{d}} + \gamma)  \psi_0(\Omega)  = E_0(\gamma) \ \Sigma(\Omega) \psi_0(\Omega) \ .
\]

Since we are interested in the limit $\gamma \rightarrow 0^+$, we can use perturbation theory and write
\[
\begin{aligned}
E_0(\gamma) &=  \sum_{k=1}^\infty  \epsilon_k \gamma^k \\
\psi_0(\Omega)  &= Y_{0,\bf{0}}(\Omega) + \sum_{k=1}^\infty \psi_0^{(k)}(\Omega) \ \gamma^k
\end{aligned}
\]
where $Y_{0,\bf{0}}(\Omega)   = 1/\sqrt{\mathrm{Vol}(S^{d})}$.

At each order in $\gamma$ one can then obtain a set of coupled differential equations, which can be solved starting from the lowest order.

\subsection*{Order $0$}
\[
-\Delta_{S^{d}}  \psi_0^{(0)}(\Omega)  = 0 \ .
\]
with
\[
\begin{aligned}
\epsilon_0 &= 0  \\
\psi_0^{(0)}(\Omega)   &=  1/\sqrt{\mathrm{Vol}(S^{d})}
\end{aligned}
\]

\subsection*{Order $1$}
\begin{equation}
-\Delta_{S^{d}}    \psi_0^{(1)}(\Omega)   +  \psi_0^{(0)}(\Omega) = \epsilon_1 \ \Sigma(\Omega) \psi_0^{(0)}(\Omega) \ .
\label{eq:pt1}
\end{equation}

By projecting this equation over the zero mode one obtains 
\[
\epsilon_1 =   \frac{\mathrm{Vol}(S^{d})}{\int   \Sigma(\Omega)  d\Omega}
\]

Because $\psi_0^{(1)}(\Omega)$ is orthogonal to $\psi_0^{(0)}(\Omega)$, it can be decomposed in the basis of hyper-spherical harmonics as 
\[
\psi_0^{(1)}(\Omega) =  \sum_{l=1}^\infty \sum_{{\bf m}}  c_{l {\bf m}} Y_{l,{\bf m}}(\Omega) \ .
\]

It is easy to verify that
\[
\psi_0^{(1)}(\Omega) = \frac{\epsilon_1}{\sqrt{\mathrm{Vol}(S^{d})}}
 \int G^{(d)}(\Omega,\Omega') \Sigma(\Omega') d\Omega'
\]
is a solution of Eq.~(\ref{eq:pt1}).

\subsection*{Order $k \geq 2$}

To order $k\geq 2$ the following equation holds
\[
-\Delta_{S^{d}}    \psi_0^{(k)}(\Omega)   +  \psi_0^{(k-1)}(\Omega) = \Sigma(\Omega)  \sum_{j=0}^{k-1} \epsilon_{j} \  \psi_0^{(k-j)}(\Omega) \hspace{0.5cm}, \hspace{0.5cm} k=2,3,\dots
\]
with solutions
\begin{subequations}
\begin{align}
\epsilon_k &=  - \frac{\sum_{j=1}^{k-1} \epsilon_j \langle \psi_0^{(0)} | \Sigma | \psi_0^{(k-j)} \rangle}{\langle \psi_0^{(0)} |  \Sigma | \psi_0^{(0)} \rangle} \label{eq_EN_recur}\\
\psi_0^{(k)}(\Omega) &=  \sum_{j=1}^k \epsilon_j \int G^{(d)}(\Omega,\Omega') \Sigma(\Omega') \psi_0^{(k-j)}(\Omega') d\Omega'  \nonumber \\
-& \int   G^{(d)}(\Omega,\Omega') \psi_0^{(k-1)}(\Omega') d\Omega'  \  .
\label{eq_WF_recur} 
\end{align}
\end{subequations}

Notice that the solutions to order $k$ are expressed in terms of all solutions of lower order and, at least in principle, one can calculate their expression to any order.

In our case we have calculated explicitly the expressions for the energy and for the eigenfunction of the zero mode up to order $\gamma^4$ (we omit to report the explicit expressions for the corrections to the  eigenfunction because we don't need them in our calculation):
\begin{equation}
\begin{aligned}
\epsilon_2  &= - \mathrm{Vol}(S^{d})^2 \frac{\int \!  \Sigma(\Omega)\, G^{(d,0)}(\Omega,\Omega')\, \Sigma(\Omega')  \ d\Omega d\Omega'}{\left(\int\Sigma(\Omega) d\Omega\right)^3} \\
\epsilon_3 &=  \mathrm{Vol}(S^{d})^2 \frac{\int \! \Sigma(\Omega) \, G^{(d,1)}(\Omega,\Omega')\, \Sigma(\Omega') d\Omega d\Omega'}{\left(\int\Sigma(\Omega) d\Omega\right)^3} \\
&+ 2  \ \mathrm{Vol}(S^{d})^3 \frac{\left( \int \! \Sigma(\Omega) \, G^{(d,0)}(\Omega,\Omega')\, \Sigma(\Omega')  d\Omega d\Omega' \right)^2}{\left(\int\Sigma(\Omega) d\Omega\right)^5}  \\
&-  \mathrm{Vol}(S^{d})^3 \frac{\int \! \Sigma(\Omega) \, G^{(d,0)}(\Omega, \Omega')\, \Sigma(\Omega') \, G^{(d,0)}(\Omega',\Omega'')\, \Sigma(\Omega'') d\Omega d\Omega' d\Omega''}{\left(\int\Sigma(\Omega) d\Omega\right)^4}  \\
\epsilon_4 &=  - \mathrm{Vol}(S^{d})^2  \frac{\int \! \Sigma(\Omega) \, G^{(d,2)}(\Omega,\Omega')\, \Sigma(\Omega') d\Omega d\Omega'}{\left( \int \Sigma(\Omega) d\Omega\right)^3} \\
&+  \mathrm{Vol}(S^{d})^3 \left[   \frac{\int  \Sigma(\Omega) \, G^{(d,0)}(\Omega,\Omega')\,\Sigma(\Omega') \, G^{(d,1)}(\Omega',\Omega'') \,\Sigma(\Omega'') d\Omega d\Omega' d\Omega''}{\left( \int \Sigma(\Omega) d\Omega\right)^4} 
\right.  \\
&- \left.  4   \frac{\left( \int  \Sigma(\Omega)\, G^{(d,0)}(\Omega,\Omega')\,\Sigma(\Omega')\,  d\Omega d\Omega'  \right)  \, \left( \int \! \Sigma(\Omega) \, G^{(d,1)}(\Omega,\Omega')\,\Sigma(\Omega') d\Omega d\Omega'  \right)}{\left( \int \Sigma(\Omega) d\Omega\right)^5} \right. \\
&+ \left.   \frac{\int \Sigma(\Omega) \, G^{(d,1)}(\Omega,\Omega')\,\Sigma(\Omega')\, G^{(d,0)}(\Omega',\Omega'') \,\Sigma(\Omega'') d\Omega d\Omega' d\Omega''}{\left( \int \Sigma(\Omega) d\Omega\right)^4} \right] \\
&+ \mathrm{Vol}(S^{d})^4 \left[  - 5   \frac{\left( \int  \Sigma(\Omega) \, G^{(d,0)}(\Omega,\Omega')\,\Sigma(\Omega') d\Omega d\Omega' \right)^3}{\left( \int \Sigma(\Omega) d\Omega\right)^7} \right. \\
&+ \left.  5   \frac{ \int \left(\Sigma(\Omega) \, G^{(d,0)}(\Omega,\Omega') \,\Sigma(\Omega') d\Omega d\Omega' \right)   \int \!\left(\Sigma(\Omega) \, G^{(d,0)}(\Omega,\Omega')\,\Sigma(\Omega')\, G^{(d,0)}(\Omega', \Omega'')\,\Sigma(\Omega'') d\Omega d\Omega' d\Omega''\right) }{\left( \int \Sigma(\Omega) d\Omega\right)^6} \right. \\
&- \left.  \frac{ \int  \left(\Sigma(\Omega) \, G^{(d,0)}(\Omega,\Omega')\,\Sigma(\Omega')\, G^{(d,0)}(\Omega',\Omega'')\,\Sigma(\Omega'')\, G^{(d,0)}(\Omega'',\Omega''')\,\Sigma(\Omega''') d\Omega d\Omega' d\Omega'' d\Omega'''\right) 
}{\left( \int \Sigma(\Omega) d\Omega\right)^5} 
\right]
\end{aligned}
\end{equation}

\section{Some integrals}
\label{appB}

We work out the integrals introduced in Section ~\ref{sec:renorm}.

We have
\begin{subequations}
\begin{align}
\mathcal{I}_1^{(q)} &\equiv \int \Sigma(\Omega) G^{(d,q)}(\Omega,\Omega') \Sigma(\Omega')  d\Omega d\Omega' \nonumber \\ 
 &=  \sum'_{l,\vec{m}} \int \frac{Y_{l\vec{m}}(\Omega) Y_{l\vec{m}}^\star(\Omega')}{(l (l+d-1))^{q+1}} 
        \left(1 + \sum'_{l_1,\vec{m}_1} c_{{l_1} \vec{m}_1} Y_{{l_1} \vec{m}_1}(\Omega)\right) \nonumber \\
&\cdot \left(1 + \sum'_{l_2,\vec{m}_2} c_{{l_2} \vec{m}_2} Y_{{l_2} \vec{m}_2}(\Omega)\right) d\Omega d\Omega' \nonumber \\
&= \sum'_{l,\vec{m}} \frac{|c_{l\vec{m}}|^2}{(l (l+d-1))^{q+1}} 
\end{align}
\end{subequations}
where the primed sum excludes the zero mode.

Similarly we can calculate the remaining integrals:
\begin{subequations}
\begin{align}
\mathcal{I}_2^{(q,p)} &= \sum'_{l,\vec{m}}  \sum'_{l',\vec{m}'} \int \frac{Y_{l\vec{m}}(\Omega) Y_{l\vec{m}}^\star(\Omega') Y_{l'\vec{m}'}(\Omega') 
	Y_{l'\vec{m}'}^\star(\Omega'')}{(l (l+d-1))^{q+1}  (l' (l'+d-1))^{p+1}} \nonumber \\
& \left(1 + \sum'_{l_1,\vec{m}_1}  c^\star_{{l_1} \vec{m}_1} Y^\star_{{l_1} \vec{m}_1}(\Omega)\right)  
\left(1 + \sum'_{l_2,\vec{m}_2} c_{{l_2} \vec{m}_2} Y_{{l_2} \vec{m}_2}(\Omega')\right) \nonumber \\
& \left(1 + \sum'_{l_3,\vec{m}_3} c_{{l_3} \vec{m}_3} Y_{{l_3} \vec{m}_3}(\Omega'')\right)
d\Omega d\Omega' d\Omega'' \nonumber \\
&= \sum'_{l,\vec{m}}  \frac{|c_{l\vec{m}}|^2}{(l (l+d-1))^{p+q+2}} \nonumber \\
&+ \sum'_{l,\vec{m}}  \sum'_{l',\vec{m}'}  \sum'_{l_2,\vec{m}_2} 
\frac{c_{l\vec{m}}^\star c_{l'\vec{m}'} c_{l_2 \vec{m}_2}}{(l (l+d-1))^{q+1} (l' (l'+d-1))^{p+1}}
W_{l,\vec{m},l',\vec{m}',l_2,\vec{m}_2} 
\end{align}
\end{subequations}
where
\begin{subequations}
\begin{align}
W_{l_1,\vec{m}_1,l_2,\vec{m}_2,l_3,\vec{m}_3} \equiv& \int Y_{l_1,\vec{m}_1}^\star(\Omega) Y_{l_2,\vec{m}_2}(\Omega) Y_{l_3,\vec{m}_3}(\Omega) d\Omega 
\end{align} \ .
\end{subequations}

\begin{subequations}
\begin{align}
\mathcal{I}_3^{(q,p,r)} &= \sum'_{l,\vec{m}} \sum'_{l',\vec{m}'} \sum'_{l'',\vec{m}''} \int \frac{Y_{l\vec{m}}(\Omega) Y_{l\vec{m}}^\star(\Omega') Y_{l'\vec{m}'}(\Omega') 
	Y_{l'\vec{m}'}^\star(\Omega'')  Y_{l''\vec{m}''}(\Omega'') Y_{l''\vec{m}''}^\star(\Omega''')}{(l (l+d-1))^{q+1}  (l' (l'+d-1))^{p+1} (l'' (l''+d-1))^{r+1}} \nonumber \\
& \left(1 + \sum'_{l_1,\vec{m}_1}  c^\star_{{l_1} \vec{m}_1} Y^\star_{{l_1} \vec{m}_1}(\Omega)\right)  
\left(1 + \sum'_{l_2,\vec{m}_2} c_{{l_2} \vec{m}_2} Y_{{l_2} \vec{m}_2}(\Omega')\right) \nonumber \\
& \left(1 + \sum'_{l_3,\vec{m}_3} c_{{l_3} \vec{m}_3} Y_{{l_3} \vec{m}_3}(\Omega'')\right)
\left(1 + \sum'_{l_4,\vec{m}_4} c_{{l_4} \vec{m}_4} Y_{{l_4} \vec{m}_4}(\Omega''')\right) d\Omega d\Omega' d\Omega'' d\Omega''' \nonumber \\
&= \sum'_{l,\vec{m}} \frac{|c_{l\vec{m}}|^2}{(l (l+d-1))^{p+q+r+3}} \nonumber \\
&+ \sum'_{l,\vec{m}} \sum'_{l',\vec{m}'} \sum'_{l_1,\vec{m}_1} 
c_{l\vec{m}}^\star c_{l'\vec{m}'} c_{l_1 \vec{m}_1} W_{l,\vec{m},l',\vec{m}',l_1,\vec{m}_1} \nonumber \\
& \cdot \left[
\frac{1}{(l (l+d-1))^{q+1} (l' (l'+d-1))^{p+r+2}}
 	+ \frac{1}{(l (l+d-1))^{q+p+2} (l' (l'+d-1))^{r+1}}	\right] \nonumber \\
 &+ \sum'_{l\vec{m}} \sum'_{l'\vec{m}'} \sum'_{l''\vec{m}''} \sum'_{l_2 \vec{m}_2} \sum'_{l_3 \vec{m}_3} \frac{c^\star_{l \vec{m}} c_{l_2 \vec{m}_2} c_{l_3 \vec{m}_3} c_{l'' \vec{m}''}}{(l (l+d-1))^{q+1} (l' (l'+d-1))^{p+1} (l'' (l''+d-1))^{r+1}} \nonumber \\ 
&\cdot W_{l,\vec{m},l',\vec{m}',l_2,\vec{m}_2} \cdot W_{l',\vec{m}',l'',\vec{m}'',l_3,\vec{m}_3} \nonumber 
\end{align}
\end{subequations}

\begin{subequations}
\begin{align}
\mathcal{J}_1^{(q, p)}  &= \sum'_{l\vec{m}} \sum'_{l'\vec{m}'} \int \frac{Y_{l\vec{m}}(\Omega) Y_{l\vec{m}}^\star(\Omega') Y_{l'\vec{m}'}(\Omega') Y_{l'\vec{m}'}^\star(\Omega)}{(l (l+d-1))^{q+1} (l' (l'+d-1))^{p+1}}  \nonumber \\
&\cdot \left(1 + \sum'_{l_1 \vec{m}_1 } c_{{l_1} \vec{m}_1} Y_{{l_1} \vec{m}_1}(\Omega')\right) 
\left(1 + \sum'_{l_2 \vec{m}_2 } c^\star_{{l_2} \vec{m}_2} Y^\star_{{l_2} \vec{m}_2}(\Omega)\right) d\Omega d\Omega' 
\nonumber \\
&=  \sum'_{l\vec{m}}  \frac{1}{(l (l+d-1))^{p+q+2}} 
+ 2 \sum'_{l\vec{m}}  \sum'_{l_1 \vec{m}_1} \frac{c_{l_1,\vec{m}_1}}{(l (l+d-1))^{p+q+2}} W_{l,\vec{m},l,\vec{m},l_1,\vec{m}_1} \nonumber \\
&+ \sum'_{l\vec{m}} \sum'_{l' \vec{m}'} \sum'_{l_1 \vec{m}_1} \sum'_{l_2 \vec{m}_2} 
\frac{c^\star_{l_2,\vec{m}_2} c_{l_1,\vec{m}_1}}{(l (l+d-1))^{q+1} (l' (l'+d-1))^{p+1}} \nonumber \\
&\cdot W_{l,\vec{m},l',\vec{m}',l_1,\vec{m}_1} W^\star_{l,\vec{m},l',\vec{m}',l_2,\vec{m}_2} 
\end{align}
\end{subequations}   
and
\begin{align}
\mathcal{J}_2^{(q, p,r)} &= \sum'_{l\vec{m}} \frac{1}{(l (l+d-1))^{p+q+r+3}} \nonumber \\
&+ 3 \sum'_{l\vec{m}} \ \sum'_{l_1 \vec{m}_1} \frac{c_{l_1 \vec{m}_1}}{(l (l+d-1))^{p+q+r+3}} W_{l,\vec{m},l,\vec{m},l_1,\vec{m}_1} \nonumber \\
&+ \sum'_{l\vec{m}} \sum'_{l'\vec{m}'}  \sum'_{l_1 \vec{m}_1} \sum'_{l_2 \vec{m}_2} c_{l_1, \vec{m}_1} c_{l_2,\vec{m}_2} W_{l',\vec{m}',l,\vec{m},l_1,\vec{m}_1} W_{l,\vec{m},l',\vec{m}',l_2,\vec{m}_2} \nonumber \\
&\quad \cdot \left[
\frac{1}{(l (l+d-1))^{q+1} (l' (l'+d-1))^{p+r+2}} \right. \nonumber \\
&\quad \quad + \frac{1}{(l (l+d-1))^{p+1} (l' (l'+d-1))^{q+r+2}} \nonumber \\
&\quad \quad + \left. \frac{1}{(l (l+d-1))^{r+1} (l' (l'+d-1))^{p+q+2}}
\right] \nonumber \\
& + \sum'_{l\vec{m}} \sum'_{l'\vec{m}'} \sum'_{l''\vec{m}''} \sum'_{l_1 \vec{m}_1} \sum'_{l_2 \vec{m}_2} \sum'_{l_3 \vec{m}_3}
\frac{c_{{l_1} \vec{m}_1} c_{{l_2} \vec{m}_2} c_{{l_3} \vec{m}_3}  }{(l (l+d-1))^{q+1} (l' (l'+d-1))^{p+1} (l'' (l''+d-1))^{r+1}} \nonumber \\
& \cdot W_{l'',\vec{m}'',l,\vec{m},l_1,\vec{m}_1} W_{l,\vec{m},l',\vec{m}',l_2,\vec{m}_2} W_{l',\vec{m}',l'',\vec{m}'',l_3,\vec{m}_3}
\end{align}

\end{appendix}

\bibliography{myrefs.bib}

@article{Leonhardt2009PerfectImaging,
  author  = {Leonhardt, Ulf},
  title   = {Perfect imaging without negative refraction},
  journal = {New Journal of Physics},
  volume  = {11},
  year    = {2009},
  pages   = {093040},
  doi     = {10.1088/1367-2630/11/9/093040}
}

@article{TrompDahlen1993PotentialRepresentation,
  author  = {Tromp, Jeroen and Dahlen, F. A.},
  title   = {Variational principles for surface wave propagation on a laterally heterogeneous Earth---III. Potential representation},
  journal = {Geophysical Journal International},
  volume  = {112},
  number  = {2},
  pages   = {195--209},
  year    = {1993},
  doi     = {10.1111/j.1365-246X.1993.tb01449.x}
}

@article{LevyLeblond1995PDM,
  author  = {L{\'e}vy-Leblond, Jean-Marc},
  title   = {Position-dependent effective mass and Galilean invariance},
  journal = {Physical Review A},
  volume  = {52},
  number  = {3},
  pages   = {1845--1849},
  year    = {1995},
  doi     = {10.1103/PhysRevA.52.1845}
}

@article{Amore20,
  author  = {Paolo Amore},
  title   = {Exact sum rules for heterogeneous spherical drums},
  journal = {Annals of Physics},
  volume  = {412},
  pages   = {168041},
  year    = {2020},
  doi     = {10.1016/j.aop.2019.168041}
}

@article{Itzykson86,
  author  = {C. Itzykson and P. Moussa and J. M. Luck},
  title   = {Sum rules for quantum billiards},
  journal = {Journal of Physics A},
  volume  = {19},
  pages   = {L111--L115},
  year    = {1986},
  doi     = {10.1088/0305-4470/19/3/004}
}

@article{Berry86,
  author  = {M. V. Berry},
  title   = {Spectral zeta functions for {A}haronov--{B}ohm quantum billiards},
  journal = {Journal of Physics A},
  volume  = {19},
  pages   = {2281--2296},
  year    = {1986},
  doi     = {10.1088/0305-4470/19/12/015}
}

@article{Steiner87,
  author  = {Frank Steiner},
  title   = {Spectral Sum Rules for the Circular {A}haronov--{B}ohm Quantum Billiard},
  journal = {Fortschritte der Physik/Progress of Physics},
  volume  = {35},
  number  = {1},
  pages   = {87--114},
  year    = {1987},
  doi     = {10.1002/prop.2190350105}
}

@article{Steiner85,
  author  = {F. Steiner},
  title   = {Magic sum rules for confinement potentials},
  journal = {Physics Letters B},
  volume  = {159},
  number  = {4--6},
  pages   = {397--402},
  year    = {1985},
  doi     = {10.1016/0370-2693(85)90276-X}
}

@article{Kvitsinsky96,
  author  = {Andrei A. Kvitsinsky},
  title   = {Zeta functions of nearly circular domains},
  journal = {Journal of Physics A: Mathematical and General},
  volume  = {29},
  number  = {19},
  pages   = {6379},
  year    = {1996},
  doi     = {10.1088/0305-4470/29/19/022}
}

@article{Dittmar02,
  author  = {Bodo Dittmar},
  title   = {Sums of reciprocal eigenvalues of the Laplacian},
  journal = {Mathematische Nachrichten},
  volume  = {237},
  number  = {1},
  pages   = {45--61},
  year    = {2002}
}

@article{Dittmar11,
  author  = {B. Dittmar and M. Hantke},
  title   = {About a Polya-Schiffer inequality},
  journal = {Annales UMCS, Mathematica},
  volume  = {65},
  number  = {2},
  pages   = {29--44},
  year    = {2011},
  doi     = {10.2478/v10062-011-0011-8}
}

@article{Dostanic11,
  author  = {M. R. Dostani\'{c}},
  title   = {Regularized trace of the inverse of the {D}irichlet Laplacian},
  journal = {Communications on Pure and Applied Mathematics},
  volume  = {64},
  pages   = {1148--1164},
  year    = {2011},
  doi     = {10.1002/cpa.20368}
}

@article{Amore13A,
  author  = {Paolo Amore},
  title   = {Exact sum rules for inhomogeneous strings},
  journal = {Annals of Physics},
  volume  = {338},
  pages   = {341--360},
  year    = {2013},
  doi     = {10.1016/j.aop.2013.05.011}
}

@article{Amore13B,
  author  = {Paolo Amore},
  title   = {Exact sum rules for inhomogeneous drums},
  journal = {Annals of Physics},
  volume  = {336},
  pages   = {223--244},
  year    = {2013},
  doi     = {10.1016/j.aop.2013.05.010}
}

@article{Amore14,
  author  = {Paolo Amore},
  title   = {Exact sum rules for inhomogeneous systems containing a zero mode},
  journal = {Annals of Physics},
  volume  = {349},
  pages   = {253--267},
  year    = {2014},
  doi     = {10.1016/j.aop.2014.06.019}
}

@article{Amore18,
  author  = {Paolo Amore},
  title   = {Exact sum rules for quantum billiards of arbitrary shape},
  journal = {Annals of Physics},
  volume  = {388},
  pages   = {12--24},
  year    = {2018},
  doi     = {10.1016/j.aop.2017.11.003}
}

@article{Avery85,
  author  = {Zhen-Yi Wen and John Avery},
  title   = {Some properties of hyperspherical harmonics},
  journal = {Journal of Mathematical Physics},
  volume  = {26},
  number  = {3},
  pages   = {396--403},
  year    = {1985},
  doi     = {10.1063/1.526621}
}

@article{Szmytkowski06,
  author  = {Rados{\l}aw Szmytkowski},
  title   = {Closed form of the generalized {G}reen's function for the {H}elmholtz operator on the two-dimensional unit sphere},
  journal = {Journal of Mathematical Physics},
  volume  = {47},
  number  = {6},
  pages   = {063506},
  year    = {2006},
  doi     = {10.1063/1.2203430}
}

@article{Szmytkowski07,
  author  = {Rados{\l}aw Szmytkowski},
  title   = {Closed forms of the {G}reen's function and the generalized {G}reen's function for the {H}elmholtz operator on the {N}-dimensional unit sphere},
  journal = {Journal of Physics A: Mathematical and Theoretical},
  volume  = {40},
  number  = {5},
  pages   = {995},
  year    = {2007},
  doi     = {10.1088/1751-8113/40/5/009}
}

@article{Amore10,
  author  = {Paolo Amore},
  title   = {Spectroscopy of drums and quantum billiards: Perturbative and nonperturbative results},
  journal = {Journal of Mathematical Physics},
  volume  = {51},
  pages   = {052105},
  year    = {2010},
  doi     = {10.1063/1.3364792}
}

@article{Bandara21,
  author  = {L. Bandara and M. Nursultanov and J. Rowlett},
  title   = {Eigenvalue asymptotics for weighted Laplace equations on rough {R}iemannian manifolds with boundary},
  journal = {Annali della Scuola Normale Superiore di Pisa},
  volume  = {22},
  number  = {4},
  pages   = {1843--1878},
  year    = {2021},
  doi     = {10.2422/2036-2145.201902_003}
}

\end{document}